\documentclass[useAMS,usenatbib]{article}
\usepackage{times}
\usepackage{amssymb}
\usepackage{amsfonts}
\usepackage{graphicx}
\newcommand{\msun}{{$M_\odot$}}

\title{The Empirical Mass-Luminosity Relation for Low Mass Stars}

\author{Fang Xia$^{1,2}$\thanks{E-mail:xf@pmo.ac.cn}, Shulin Ren$^{1,2}$
and Yanning Fu$^{1}$\\
 $^{1}$Purple Mountain Observatory, Chinese
Academy of Sciences, Nanjing 210008, China\\
$^{2}$Graduate School of Chinese Academy of Sciences, Beijing
100039, China}
\begin{document}
\maketitle
\begin{abstract}
This work is devoted to improving  empirical mass-luminosity
relations (MLR) and mass-metallicity-luminosity relation (MMLR)  for
low mass stars. For these stars, observational data in the
mass-luminosity plane or the mass-metallicity-luminosity space
subject to non-negligible errors in all coordinates with different
dimensions. Thus a reasonable weight assigning scheme is needed for
obtaining more reliable results. Such a scheme is developed, with
which each data point can have its own due contribution. Previous
studies have shown that there exists a plateau feature in the MLR.
Taking into account the constraints from the observational
 luminosity function, we find
by fitting the observational data using our weight assigning scheme
that the plateau spans from 0.28 \msun to 0.50 \msun.
Three-piecewise continuous improved MLRs in K, J, H and V bands,
respectively, are obtained. The visual MMLR is also improved based
on our K band MLR and the available observational metallicity data.
\end{abstract}

 \noindent {\it Keywords:} {\ stars: low-mass--stars:
fundamental parameters--methods: numerical}
\section{Introduction}

Mass is one of the most fundamental parameters of a star.
Unfortunately, stellar mass is difficult to be determined directly.
Therefore, it is often estimated by mass-luminosity relation (MLR).

Since the pioneering papers of Hertzsprung (1923) and Russell et al.
(1923) there have been many studies devoted to improving and
understanding MLR. To date, the relation has been well constrained
for solar-type and intermediate stars \cite{Del00}. But the
empirical MLR remains poorly defined for very low mass stars due to
scarcity of data. On the other hand, however, the empirical MLR for
very low mass stars is crucial in many aspects of astronomy and
astrophysics. For example, without an accurate MLR in this mass
interval, the total luminous mass of the Galaxy will never be well
determined \cite{Henry98}. This is because very low mass stars
account for at least 70\% of all stars, and they make up more than
40\% of the total stellar mass of the Galaxy \cite{Henry99}.

Recently, the MLR for very low mass stars has been improved
significantly, thanks to the painstaking observations and the
required orbital analysis. Henry et al. (1999) obtained or improved
some dynamical masses of very low mass stars and discussed the MLR
for the mass interval 0.2 \msun to 0.08 \msun. Delfosse et al.
(2000) pointed out that MLR is tight only in near-infrared bands.
Together with the improved MLRs in K, J and H bands, they suggested
that the large dispersion in V band should be due to the difference
in stellar metallicity. In Bonfils et al. (2005), with quantitative
metallicity estimations, the authors verified this metallicity
dependency and provided visual mass-metallicity-luminosity relation
(MMLR) for the first time.

Despite of these progresses, the empirical MLR for low mass stars
obtained so far is still not satisfying. The existence of a
transitional mass interval, where the derivative of MLR presents a
plateau feature, is an intrinsic obstacle to a satisfying result.
This is because a continuous yet segmented model is needed for MLR.
While more data are often necessary for fitting to a segmented model
than a non-segmented one, some data with relatively low accuracy
were discarded in the previous work. This is not without reason, as
many of the data are prone to systematics. Obviously, to remove such
data, it should be more reasonable to require that the remaining
data set is well behaved. And at the same time, one should avoid
discarding data more than necessary, such that the original data set
can be fully utilized. Based on a newly developed weight assigning
scheme, we are able to achieve this by an iterative process and, as
thus, give a more confident MLR. In comparison with previous work
\cite{Del00,HM93}, the improvement of MLR in (0.7 \msun, 1.0\msun)
also comes from incorporating more observational data, and that in
(0.1\msun, 0.7\msun) from taking into account the underlying
physics.

In section 2 and section 3, respectively, we describe our sample
data and develop a fitting method with a reasonable weight assigning
scheme. In section 4, the best-fitting three-piecewise empirical
MLRs in K, J, H and V bands, the MMLR in V band are  provided.
Concluding remarks are given in the last section.

\section{DATA COLLECTION}

In order to map out MLR for low mass main sequence stars, we need a
sample of dynamical masses derived from orbital analysis
\cite{Henry04} and absolute magnitudes. By searching the
literatures, the resulting sample of 48 main sequence stars  is
listed in Table 1. The columns of Table 1 are, respectively, name of
these stars, the dynamical mass ($M\pm \Delta M$) (spanning from
0.07 \msun $\sim$1.086 \msun), absolute V magnitude ($M_V\pm\Delta
M_{V}$), absolute K magnitude ($M_K\pm \Delta M_{K}$), absolute J
magnitude ($M_J\pm\Delta M_{J}$) and absolute H magnitude
($M_H\pm\Delta M_{H}$). Most of the data can be found directly from
the literatures except that some values of $M_V, M_J$ and $M_H$ are
obtained from apparent magnitudes (color index if necessary) and
parallaxes. The types of spectrum as well as references are also
indicated.

\begin{table*}
  \begin{minipage}{180mm}
  \caption{Masses, absolute magnitudes and spectrums of the sample stars.  The data in italic are
discarded in fitting MLR.}
\begin{tabular}{@{}lllrlllrlrlll@{}}
  \hline \hline
Name &$M $ & $\Delta{M}$ &$M_V$& $\Delta{M_V}$ &$M_K$ &
$\Delta{M_K}$ &$M_J$& $\Delta{M_J}$&$M_H$ & $\Delta{M_H}$&Spectrum&
Reference\\
& (\msun) &(\msun) & (mag) & (mag) & (mag)& (mag)& (mag)& (mag)&(mag) & (mag)& &\\
 \hline
GL22A&0.43&0.039&10.56 & 0.07&6.44&0.10& & & 6.74  &0.08&M2V  &1\&2\&9\\
GL22C     & 0.14 & 0.014&13.64&  0.12 &8.43 & 0.13 &&&8.85&  0.10& M2V &1\&2\&9\\
GL25A &0.94 &0.088& 4.67 &0.088&3.82&0.27&\emph{3.18} & \emph{0.08}  &3.88 & 0.26&G8V  & 6\&9\&10$^{a}$\\
GL25B &0.7&0.079&4.99& 0.088&3.98&0.27 &&&4.13&  0.26&G8V  & 6\&9\&10$^{a}$\\
GL65A   &0.102 &0.01 &15.41 & 0.05&8.76 & 0.07 &\emph{9.68} & \emph{0.05}&  \emph{9.15} & \emph{0.03}& M5.5V &8\\
GL65B &0.100 &0.01 &15.87  &0.06&9.16  &0.07&10.06 & 0.05  &9.45  &0.03 &M6V  &8\\
GL67A &0.933 &0.231& 4.45 &0.03&\emph{2.87} & \emph{0.12} &\emph{3.14} & \emph{0.12} & 2.90 & 0.12&M4V&9\\
GL67B  & 0.28 &0.071&  12.07& 0.5&7.30  &0.13&\emph{7.52}  &\emph{0.27}  &7.40 & 0.17& M4V &9\\
GL166C  &0.177& 0.029 &12.68 & 0.03& 7.58 & 0.07&8.49 & 0.07 & \emph{7.87}  &\emph{0.07}& M4.5V &1\&9\\
GL234A     &0.2027 &0.0106 &13.07&  0.05 &7.64 & 0.04&8.52  &0.06  &7.93  &0.04&M4.5V &8\\
GL234B &0.1034& 0.0035  &16.16  &0.07 & 9.26 & 0.04&10.31 & 0.25 & 9.56 & 0.10 & M4.5V &8\\
GL340A &0.657 &0.101 & &&4.25 & 0.12& \emph{4.73} & \emph{0.12}& \emph{4.23}  &\emph{0.12}&K3V&9\\
GL340B&\emph{0.590} &\emph{0.090} &&& \emph{4.33 }& \emph{0.12}&\emph{5.08} & \emph{0.12}  &\emph{4.65} & \emph{0.12}&K3V&9\\
GL352A&\emph{0.195} &\emph{0.060}&\emph{10.80}  &\emph{0.20}&\emph{6.38}&\emph{0.19} &\emph{7.13} & \emph{0.19}  &\emph{6.59} & \emph{0.19}&M3V&9\\
GL352B&\emph{0.203}&\emph{0.062}&\emph{11.20}&  \emph{0.18}&\emph{6.61}&\emph{0.19}& \emph{7.46} & \emph{0.19}&\emph{6.85} &\emph{0.19}&M3V&9\\
GL473A &0.143 &0.011&15.01 &0.07&8.40 &0.06& 9.44 & 0.06 & 8.84 & 0.06&M5.5V  &8\\
GL473B   &0.131 &0.01 &15.00 &0.07&8.84 &0.08 &9.57 & 0.06 & 9.04&  0.07&M5.5V&8\\
GL508A &0.709 &0.179 & 8.43  &0.17&4.95 & 0.17& 5.72&  0.17&  5.12 & 0.17&M0.5V&9\\
GL508B &0.497 &0.126  &10.14  &0.19&5.87 & 0.17&6.80 & 0.18 & 6.10&  0.17&M1V&9\\
GL559A &1.086 & 0.025&   4.71& 0.05& 2.90&  0.05& 3.24& 0.05 &3.00& 0.05&G2V&9\\
GL559B &0.903&0.021&6.08 &0.05&3.78  &0.05&4.37&  0.05 & 3.89  &0.05&K1V&9\\
GL570B&0.5656& 0.0029 &9.28& 0.09 & 5.39  &0.03 &6.21&  0.03&  5.61&  0.03&M1V&6\&8$^{b}$\\
GL570C &0.377 &0.0018 &11.09 & 0.17& 6.57 & 0.04&7.40&  0.04  &6.76  &0.04&M3V&6\&8$^{b}$\\
GL623A &0.3432 &0.0301 & 10.74 & 0.05&6.46  &0.04&7.19&  0.04  &6.70 & 0.04& M2.5V&8\\
GL623B &0.1142 &0.0083  &16.02&  0.11&9.33  &0.14  & 10.47  &0.29  &9.35&  0.05&M2.5V&8\\
GL644A& 0.4155 &0.0057  &10.76&  0.06&6.35  &0.04&&&6.61&  0.05&M3V&8\\
GL661A &0.379& 0.035 &11.10 & 0.06&6.36 & 0.05 &7.10&  0.05  &6.56  &0.04 &  M4.5V &8\\
GL661B &0.369& 0.035&11.15 & 0.06& 6.78 & 0.05 &7.51 & 0.04 & 7.02&  0.04 &M4.5V&8\\
GL677A &\emph{0.285}&\emph{0.092}&\emph{8.37} & \emph{0.22}&\emph{4.94}&\emph{0.22}&\emph{5.65}  &\emph{0.21} & \emph{5.10}& \emph{0.21}&K3V&9\\
GL677B&\emph{0.251}&\emph{0.081}&\emph{8.93} & \emph{0.22} &\emph{5.22}&\emph{0.22}&\emph{5.95} & \emph{0.21} &\emph{5.41}  &\emph{0.21} &K3V&9\\
GL702A&0.9&0.074&5.63 &0.047&3.293&0.048&4.15  &0.06& 3.39&  0.05&K0V&4\&6\&9$^{b}$\\
GL702B&0.78&0.04&7.43 &0.047&4.53 & 0.04&5.63 & 0.05&&&K0V&6\&8$^{b}$\\
GL704A&0.932&  0.187&   3.95 & 0.15  &\emph{2.38}&  \emph{0.17} &\emph{2.62}&\emph{0.16}&\emph{2.47}&\emph{0.15}&F6V&9\\
GL704B &0.616 &0.124 &7.31  &0.15& 4.44 & 0.18&\emph{5.15} & \emph{0.16}&  4.87&  0.16&K4V&9\\
GL725A&0.37 & 0.061 &11.18 & 0.05& 6.72  &0.03 &7.48&  0.03 & 6.95 & 0.03&M3V&9\\
GL725B &0.316 &0.052 & 11.96 & 0.05&7.25  &0.03 &8.00 & 0.03 & 7.48 & 0.03&M3V&9\\
GL747A  &0.2137&0.0009&12.30&  0.06&7.53&0.04&&&&&M3V &8 \\
GL747B &0.1997&0.0008&12.52 & 0.06&7.63&0.04&&&&&M3V &8 \\
GL820A &0.562 &0.08 & 7.58  &0.04& 4.598 &0.015&5.32&  0.02&  4.71  &0.02&K5V&3\&4\\
GL820B &0.562 &0.08 & 8.41 & 0.04 & 4.958 &0.018&5.75 & 0.02  &5.11 & 0.02&K5V&3\&4\\
GL831A & 0.2913&  0.0125&  12.52 & 0.06&7.08 &  0.05 &&&7.36  &0.05&M4.5V &8\\
GL831B&0.1621  &0.0065 &14.62&  0.08&8.36   &0.05 &&&8.62  &0.05&M4.5V &8\\
GL860A  &0.2711 &0.0100&11.76 & 0.05& 6.95 & 0.04 & 7.84  &0.04 & 7.26 & 0.04&M3V&8\\
GL860B &0.1762 &0.0066 &13.46 & 0.09&8.32  &0.07 &9.03&  0.08  &8.40 & 0.05&M5.5V &8\\
GL866A&0.1187 &0.0011 &15.39&  0.07 &9.11  &0.32  &&&&& M5.5V&5\&7\\
GL866B&0.1145& 0.0012 &15.64 & 0.08&8.98 & 0.05 &&&9.29 & 0.04&M5.5V &5\&7\\
GL1245A   &0.11&  0.021&15.18  &0.03& 8.96 &0.03 & 9.79&  0.03  &9.32 & 0.03&M5.5V &1\&2\&9\\
GL1245C  &0.07&  0.013 &\emph{18.47}&  \emph{0.06}&\emph{9.99}  &\emph{0.04}&\emph{11.05}&  \emph{0.04}& 10.40 & 0.04& M5.5V&1\&2\&9\\
\hline
\end{tabular}

\medskip
 1.\cite{Henry99}; 2. \cite{Henry04}; 3.
\cite{Cester83}; 4. \cite{Alonso94}; 5. \cite{Woitas03};
6.\cite{Pourbaix00}; 7. \cite{Woitas00} 8. \cite{Del00}; 9.
\cite{HM93}; 10. \cite{Soderhjelm99}.\\ $^a$ parallax from 6 and
color index from 10;\ \ \ \ $^b$ parallax and apparent magnitude
from 6.

\end{minipage}
\end{table*}

In this context, it should be mentioned that accuracy is sometimes
used as a cutoff principle in deciding which star can be used for
defining MLR. For example, 10\% mass uncertainty is used as the
cutoff accuracy in Delfosse et al. (2000).  To do so, one can not
only avoid the difficulty of assigning weight but also reduce the
influence of observational systematic errors. But   in order to make
the best use of the present observations, we will not discard data
just for their relatively low accuracy (for details, see Section 4).
\section{Weight assigning scheme}

MLR is often modeled as
\begin{equation}
\log M = f(M_P),
\end{equation}
where $M_P$ is absolute magnitude, $f$ is a linear function or
polynomial. In this section, $f$ is considered as a fourth order
polynomial \cite{DK91,Del00}. As both $M$ and $M_P$ are subject to
non-negligible errors, the weight of each data point is not easy to
be assigned. Because the error of magnitude is relatively small,
there is an easy fitting method (hereafter Method 1) in which the
weight is assigned only considering the error of mass. But we found
it is not acceptable for quantitatively describing the MLR for low
mass stars. For our purpose, a new method (hereafter Method 2) in
which errors in both coordinates are used to assign appropriate
weight is developed.

To introduce Method 2, let's write
\begin{equation}
y(x)=f(M_P)=a_1+a_2x+a_3x^{2}+a_4x^{3}+a_5x^{4}
\end{equation}
and suppose ($x_i\pm\sigma_{x_i},y_i\pm\sigma_{y_i}$) (i=1,\ldots,N)
to be the observational data with respective errors. The merit
function is, as usual,
\begin{equation}
\chi^{2}=\sum_{i=1}^N[\frac{y_i-y(x_i)}{\sigma_i}]^{2},
\end{equation}
where  the reciprocal of $\sigma_i$ is the assigned weight $w_i$ of
($x_i,y_i$).

For linear model
\begin{equation}
y=a+bx,
\end{equation}
the weight of each data point can be quantified as \cite{Press92}:
\begin{equation}
w_i\equiv\frac{1}{\sigma_i}=\frac{1}{\sqrt{b^{2}\sigma_{x_i}^{2}+\sigma_{y_i}^{2}}}.
\end{equation}
In nonlinear case, the weight of each data point depends on the
local property of the fitting result, so an iteration process is
unavoidable. The main steps of the iteration process are described
as follows.

(1) A linear fitting with errors in both coordinates is performed
with the help of Press et al. (1992).

(2) With the slope ``b''  of the linear model obtained in the first
step, the weight is approximately quantified by using equation (5)
and then, the data are fitted to the given model, $y=y(x)$.

(3) By using the preliminary fitting result, the weight of
($x_i,y_i$) can be re-quantified according to the local slope. In
terms of normalized local tangent vector, the weight writes:
\begin{equation}
w_i\equiv\frac{1}{\sigma
_i/\sqrt{b^{2}_i+1}}=\sqrt{\frac{b_i^{2}+1}{b^{2}_i\sigma_{x_i}^{2}+\sigma_{y_i}^{2}}},
\end{equation}
where $b_i=f'(x_i)$.  And the model parameters are updated
accordingly.

(4) The third step is iterated until each model parameter changes
less than, say 1\%.

In order to compare the above mentioned two methods, we do the
following tests. We start with a fourth order polynomial $Log
M=f(M_V)$, named as ``true MLR'', qualitatively resembling the MLR
for very low mass stars (M $< 0.25$ \msun). We then generate
``observational data'' from the ``true MLR'' by adding Gaussian
errors at a level of the relevant observations.

There are three kinds of fitted MLRs. For each kind, 100 MLRs are
obtained with different sets of generated observational data. The
MLRs of the first and the second kinds are obtained by fitting 20
data points (approximately  the number of very low mass sample
stars) by Method 1 and Method 2, respectively. And the MLRs of the
third kind are obtained by fitting different sets of 40 data points,
using Method 2. The third kind is included because comparing results
obtained from different number of data points can shed some light on
the role of future observations in improving MLR.

Figure  1 shows typical cases of the three kinds of fitting. The
``true MLR'' and the ``observational data'' are shown as solid curve
and points with error bars, respectively. Comparing the upper two
panels, we see clearly that Method 2 is  better than Method 1. And,
as is evident in the top panel, Method 1 is apt to bring spurious
characteristics to the obtained MLR. This suggests that an improper
way of weight assigning tends to lead not only quantitative but also
qualitative difference between a fitted and a realistic MLR. From
the middle panel, we know that, by using Method 2, 20 points already
have the opportunity to be used to recover rather satisfactorily the
``true MLR''. As expected, the improvement brought by doubling
observational data is evident, as can be seen by comparing the lower
two panels. This shows the importance of future observations in
improving further the empirical MLR.
\begin{figure}[!htb]
\begin{center}
  \includegraphics[width=0.45\textwidth]{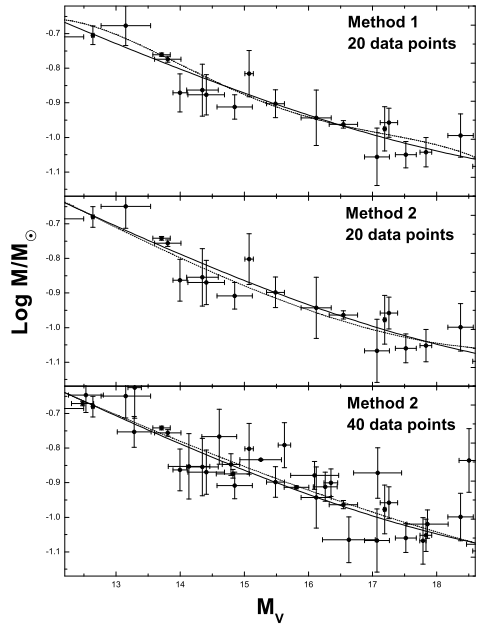}
  \end{center}
 \caption{Typical results of the three different kinds
of fitted  mass-luminosity relation for very low mass stars. In each
panel, the dots with error bars are the ``observational'' data, the
dashed and the solid lines, respectively, the fitted and the
``true'' mass-luminosity relations.} \label{fig_1}
\end{figure}

For comparing the two methods statistically, we calculate the
absolute errors ($\Delta M$) of the  masses estimated from the above
MLRs obtained using 20 data points. For each fitted MLR, $\Delta M$
at 100 randomly chosen magnitudes are calculated. In total, there
are 20000 such errors \{$\Delta M_{kij},k=1,2; i=1,\ldots,100;
j=1,\ldots,100$\}, where the subscripts $k$, $i$ and $j$ indicate
the used methods, the generated data sets and the chosen magnitudes,
respectively. The maximal (mean) $\Delta M_{2ij}$ over both $i$ and
$j$ is about 40\% (76\%) of that of $\Delta M_{1ij}$. These results
clearly show that Method 2 is better than Method 1.

The introduced weight assigning scheme is still effective in
multiple dimension fittings. For fitting functions like $z=z(x,y)$
the weight writes
\begin{equation}
w_i=\sqrt{\frac{a_i^{2}+b_i^{2}+1}{a^{2}_i\sigma_{x_i}^{2}+b^{2}_i\sigma_{y_i}^{2}+\sigma_{z_i}^{2}}},
\end{equation}
where $a_i=z'(x_i),b_i=z'(y_i)$. Tests show that much better results
can be derived when the weight of each data point is reasonably
assigned.

\section{Piecewise MLR}

Early empirical MLRs were often given in the form of a power law.
Thereafter, in order to more accord with observations,
three-piecewise continuous model was used \cite{HM93}. This is also
physically justified. In fact, as the decreasing stellar mass
reaches about 0.5 \msun, convection becomes more and more efficient
and more H$_2$ forms in the stellar atmosphere
\cite{Cester83,Laughlin97}. And when stellar mass reaches about
0.3\msun, stars should be fully convective
\cite{Copeland70,Grossman74}. As a result, the derivative of MLR
should display a plateau feature between $\sim 0.3$ \msun and $\sim
0.5$ \msun. This feature is evident from observations. However, the
observations show that the plateau feature in the MLR spans slightly
larger mass interval than the theoretically estimated one.

In fact, there is a well-known relation between mass function (MF),
luminosity function (LF) and MLR, written in V band as
\begin{equation}
LF_V= MF \times |\frac{dM}{dM_{V}}|.
\end{equation}
The term $|dM/dM_{V}|$ is the derivative of V band MLR. The golden
rule about the three quantities is:`` if an LF presents features
like peaks or dips, they should not be automatically attributed to
features of the MF, before one has excluded the possibility that
they correspond (even if not exactly) to expected features of the
MLR'' \cite{Antona98}. Synthetically considering the available
observational LFs \cite{DM83,Wielen83,Scalo86,Gould96,Reid02}, we
find that it is safe to assume $0.17\leq a\leq0.30$ and $0.50\leq
b\leq0.60$, where
 $(a,b)$ is the mass interval spanned by the plateau feature. Let $D$ be the
above-mentioned area in the $(a,b)$-plane. We note that Henry \&
McCarthy (1993) claimed according to their sampling mass and
luminosity data $(a,b)=(0.18, 0.5)$ in D. In the following, we will
improve this result, as well as the associated MLRs in various
bands, in a more sophisticated and reliable way.

We choose a three-piecewise continuous model for the MLR, where
linear model is used for the mass interval associated to the
plateau, and two-order polynomial model for the other two
\cite{HM93,Henry99}. And, with a step of 0.01 \msun, the attempted
precision for mass, we construct a set of two-dimensional grid
points on the region $D$. Each grid point corresponds to a possible
mass interval associated to the plateau. By fitting K, H, J and V
band data, respectively, there are four $\chi^{2}$ of best fitting
at each grid point. The searched $a,b$ then correspond to the grid
point with the smallest sum of these four $\chi^{2}$.

In order to remove the data point prone to systematics, an iterative
process is used. Firstly, using all the collected data, we map out
MLRs in K, J, H and V bands. Though there are some data prone to
systematics, the resulting MLRs are good first approximation. This
is because the fact that most of these data have low accuracy and
our weight assigning scheme effectively reduces their negative
influence. Therefore, as a second step, the deviation of
observational data to an MLR, approximated by the ones obtained in
the first step, can be reasonably estimated. If the absolute
expectation of this deviation is lager than 0.01, the data point
corresponding to the maximum of absolute deviation is discarded. And
the reduced data set is processed again through the two steps
described above until the absolute expectation is less than 0.01.
The discarded data are shown in italic in Table 1 and as open
circles in Fig 2. It is evident that these data are indeed deviate
from the final MLR.

 The resulting K, H, J and V band MLRs are, respectively,
\begin{displaymath}
\log M =\left \{\begin{array}{ll}
0.427-0.123M_K-0.000327M_K^{2} & \textrm{if $0.50<M<1.086$}\\
0.825-0.192M_K & \textrm{if $0.28\leq M\leq0.50$}\\
3.60\:\:-0.893M_K+0.0433\:M_K^{2} & \textrm{if $0.1<M<0.28$}
\end{array}\right.
\end{displaymath}
\begin{displaymath}
\log M =\left \{\begin{array}{ll}
0.451-0.132M_H+0.00142M_H^{2} &  \textrm{if $0.50<M<1.086$}\\
0.865-0.191M_H & \textrm{if $0.28\leq M\leq0.50$}\\
1.87\:-0.430M_H+0.0139M_H^{2} &  \textrm{if $0.1<M<0.28$}
\end{array}\right.
\end{displaymath}
\begin{displaymath}
\log M =\left \{\begin{array}{ll}
0.184-0.000028M_J-0.0111M_J^{2} &  \textrm{if $0.50<M<1.086$}\\
0.958-0.189M_J & \textrm{if $0.28\leq M\leq0.50$}\\
3.76\:-0.834M_J\:\:\:\:+0.0367M_J^{2} & \textrm{if $0.1<M<0.28$}
\end{array}\right.
\end{displaymath}
\begin{displaymath}
\log M =\left \{\begin{array}{ll}
0.213-0.0250M_V-0.00275\:M_V^{2} & \textrm{if $0.50<M<1.086$}\\
0.982-0.128M_V & \textrm{if $0.28\leq M\leq0.50$}\\
4.77\:\:-0.714M_V\:\:+0.0224M_V^{2} & \textrm{if $0.1<M<0.28$}
\end{array}\right.
\end{displaymath}

We show these MLRs in Fig 2 as solid curves. For comparison, we also
plot the results of Henry \& McCarthy (1993) (hereafter HM93) and
Delfosse et al. (2000) (hereafter Del00) as dashed lines and dotted
lines, respectively. Because Del00 improved the MLR in (0.1\msun,
0.7\msun), only the MLR in (0.7\msun, 1.0\msun) of HM93 is shown.
The gray points with error bars are the data used in fitting MLR.
\begin{figure}[!htb]
\begin{center}
\includegraphics[width=0.45\textwidth]{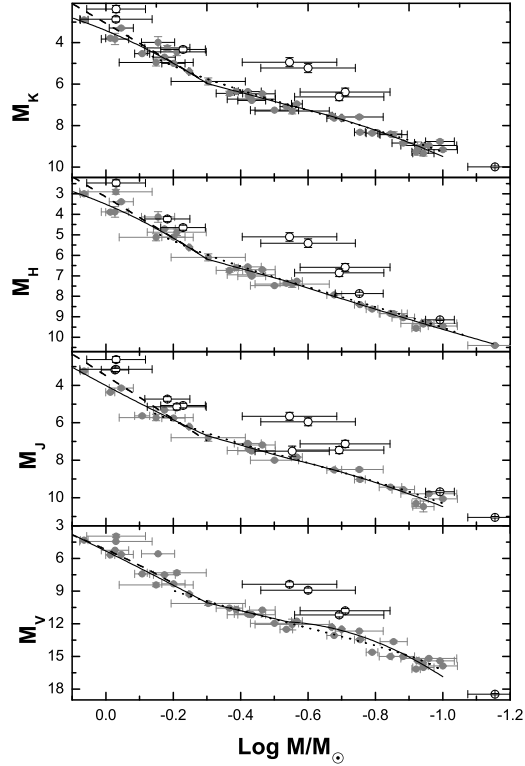}
\end{center}
\caption{Quantitative descriptions of K,  H , J and V band
mass-luminosity relations
 for low mass stars. The gray points with error bars are the data used in fitting MLR while the open circles are the discarded data,
 the solid curves show our fitted three-piecewise mass-luminosity relations, the
 dashed and the dotted lines are the results of Henry \& McCarthy (1993) and Delfosse et al.(2000), respectively.}
\label{fig_3}
\end{figure}

In comparison with HM93, we notice that our plateau region is more
accord with the theoretical result. And From Fig 2, it's easy to see
our K, J and H band MLRs are lower than HM93. Because one more
parameter is added to our MLRs, no definitive conclusion can be made
through comparing $\chi^{2}$ directly. Therefore, F-test is used to
evaluate whether adding parameters is statistically significant.

Following Lucy \& Sweeney (1971) and Pourbaix \& Jorissen (2000),
comparing the old and new models, let $L$ be the number of
parameters, $\chi^{2}$ be the weighted sum of the squares of the
residuals, $N$ be the number of observational data, the efficacy of
adding parameters can be measured by the ratio:
\begin{equation}
F =
\frac{N-L_{new}}{L_{new}-L_{old}}\cdot\frac{\chi^{2}_{old}-\chi^{2}_{new}}{\chi^{2}_{new}}.
\end{equation}
 If the hypothesis that the old model is acceptable, $F$ is distributed as $F(L_{new}-L_{old},
 N-L_{new}$) \cite{BR1992}
 and the probability density function is:
 \begin{equation}
\Phi(F)=(1+\frac{F}{\beta})^{-(1+\beta)},
\end{equation}
where $\beta=(N-L_{new})/(L_{new}-L_{old})$.  If equation (9) gives
$F=\overline{F}$, then  the probability of $F$ could have exceeded
$\overline{F}$ is:
 \begin{equation}
p=(1+\overline{F}/\beta)^{-\beta}.
\end{equation}
In fact, $p$ is the first risk error of rejecting the null
hypothesis while it is actually true. Usually 5\% level of
significance is adopted \cite{pb03}, that is, if the calculated $p$
is smaller than 0.05, the new model  can  be definitely accepted.

Comparing with HM93, one parameter is added in near-infrared MLRs.
According to equation (11), the calculated $p_K, p_J$ and $p_H$ (the
subscripts K, J and H stand for the corresponding band) are all
smaller than 0.05, which show the statistically significant
improvement to HM93. Moreover, comparing the $\chi^{2}$ taking into
account the different degrees of freedom, our $\chi^{2}$ are less
than 50\% of the corresponding value of HM93. These comparison
results show our appreciable progress in constraining MLR. We list
the values of $p$, $\chi^{2}/{(N-L)}$ in the second column of Table
2.
\begin{table}
\centering \caption{The comparison results}
\begin{tabular}{|l|l|l|l|l|}
\hline  &\multicolumn{2}{|c|}{\mbox{1.0 $\sim$ 0.7}}&
\multicolumn{2}{|c|}{\mbox{0.7$\sim$
0.07}}\\
\hline
&this work& HM93&this work&Del00\\
\hline
$p_K(\%)$&\multicolumn{2}{|c|}{\mbox{0.89}}&\multicolumn{2}{|c|}{\mbox{0.74}}\\
\hline
$\frac{\chi^{2}_K}{N-L}$&2.76&7.21&2.27&3.52\\
\hline
$p_H(\%)$&\multicolumn{2}{|c|}{\mbox{0.96}}&\multicolumn{2}{|c|}{\mbox{1.08}}\\
\hline $\frac{\chi^{2}_H}{N-L}$&2.59&6.63&1.29&2.13\\
\hline$p_J(\%)$&\multicolumn{2}{|c|}{\mbox{4.37}}&\multicolumn{2}{|c|}{\mbox{20}}\\
\hline $\frac{\chi^{2}_J}{N-L}$&4.56&14.57&1.84&2.17\\
\hline $p_V(\%)$&\multicolumn{2}{|c|}{\mbox{\ldots}}&\multicolumn{2}{|c|}{\mbox{0.02}}\\
\hline $\frac{\chi^{2}_V}{N-L}$&2.23&2.26&4.91&11.63\\
\hline
\end{tabular}
\end{table}

Del00 used a non-segmented fourth-order polynomial model in
(0.1\msun, 0.7\msun). So for his MLR, the slope in the plateau is
increased and in the other mass regions the slopes are decreased. On
the basis of an increase of three parameters, the values of  $p$ are
calculated and listed in the third column of Table 2. From this
table,  we can see that our MLRs in K, H and V bands have
statistically significant improvement to Del00. And the values of
$\chi^{2}/(N-L)$ are reduced about 40\%. For J band, due to the
number of observational data is the fewest, our improvement is not
as obvious as in the other bands and the $\chi^{2}/(N-L)$ is reduced
about 15\%.

For V band,  the difference in stellar metallicity induces a large
scatter while the near-infrared relations are much tighter
\cite{Del00}. With observational estimations of metallicity, Bonfils
et al. (2005) provided MMLR for the first time. We collect the
latest data from Woolf \& Wallerstein (2006) and with an improved
three-piecewise K band MLR, an improved V-band MMLR is provided in
the same way as Bonfils et al. (2005). The resulting formula writes:
\begin{eqnarray}
M_V
\mskip-7mu=\mskip-7mu13.4\mskip-7mu+\mskip-7mu1.29M\mskip-7mu-\mskip-7mu22.3M^{2}\mskip-7mu+\mskip-7mu12.8M^{3}\mskip-7mu+\mskip-7mu1.26[Fe/H]
\end{eqnarray}
for M $\epsilon[0.13,1.19]$\msun, [Fe/H] $\epsilon[-1.5,0.3]$.

\section{Concluding remarks}
In this paper, based on a new reasonable weight assigning scheme, we
try to make the best use of the present observations to restrict
MLRs in K, J, H, V  bands  and MMLR in V band for low mass stars.
Further improvements of this relation would have to come from future
observations \cite{Henry04}. This is especially true for the
transitional mass interval. Apart from its importance in estimating
stellar masses, the empirical MLR in this mass interval is important
in testifying simulations aimed at understanding the underlying
physics. Since the precision of observations depends heavily on
vastly different situations of individual stars, we believe the
weight assigning scheme developed in the present paper will be
helpful in the relevant studies. Furthermore, for relations like
MLR, only good schemes can demonstrate those not obvious features
thus can improve the corresponding physical investigation.

It should be pointed that, while it is very interesting to quantify
how empirical MLR in V band depends on metallicity, further
improvements to the presently available observations seem to be
necessary. Besides, the age-dependency of empirical MLR, which
remains to be poorly defined due to the fact that there are few
confident observations available \cite{Henry04}, also demands more
painstaking observational efforts.

\noindent {\it acknowledgments:} {\ We thank the reviewer whose
critical comments help us a lot in improving the manuscript. We are
also grateful to Dr. Todd J. Henry and Dr. X. Delfosse for their
helps. This work is supported by NSFC 10473025 and NSFC 10233020.}

\clearpage
\end{document}